# The Solar Wind and Climate: Evaluating the Influence of the Solar Wind on Temperature and Teleconnection Patterns Using Correlation Maps

Kiminori Itoh*, Shinya Matsuo#, Hiroshi Serizawa, Kazuyoshi Yamashita and Takashi Amemiya

Graduate School of Environment and Information Sciences, Yokohama National University, 79-7 Tokiwadai, Hodogaya-ku, Yokohama 240-8501, Japan.
# Current Address: Hino Systech Corporation, 1-20-2 Minamirokugo, Ota-ku, Tokyo 144-0045, Japan
*Corresponding author: itohkimi@ynu.ac.jp

**Abstract**

Evaluating the magnitude of natural climate variations is important because it can greatly affect future climate policies. As an example, we examine the influence of changes in solar activity (solar wind in particular) on surface temperatures ($T_s$) and major teleconnection patterns such as the Arctic Oscillation and Pacific Decadal Oscillation. We compared correlation maps (spatial distribution of correlation coefficient) for a combination of $T_s$ and a geomagnetic index (*aa*, an indicator of solar wind strength) and a combination of $T_s$ and the teleconnection patterns. The phase of the quasi-biennial oscillation of the equatorial zonal wind and magnitude of sunspot number were considered. As a result, we found that the influence of the solar wind is as strong as that of the teleconnection patterns and hence, the former appears to affect the climate via the latter. It was also found that both the solar wind and ultraviolet change should be considered to explain the influence of solar activity variability, i.e., a multi-pathway scheme is necessary.

## 1. Introduction

Upon formulating effective climate policies, the influences of various anthropogenic and natural climate forcings should be taken into account [1]. The former includes greenhouse gases [2], aerosols [3–5], and land-use change [6–8]. Pielke et al. [1] emphasized the significance of a diverse range of these anthropogenic factors with respect to mitigation and adaptation policies, as well as the importance of natural variations.

In this respect, we recently showed that climate policies had to be greatly altered when natural factors proved significant, using the scenario planning method [9, 10]. In fact, variations of these factors (internal ones such as oceans and clouds, as well as external ones such as the sun and moon) and their influences on climate and regional/local societies are potentially much greater than previously thought (as shown by several examples below). These factors are not necessarily independent, i.e., oceans and clouds can be affected by the sun and moon, and teleconnection patterns (TPs) may be interrelated.

**Ocean;** Levitus et al. [11] found multidecadal temperature oscillations with magnitude as large as

4 °C for the Barents Sea at depths of 100–150 m. Timing of the oscillation coincided with the Atlantic Multidecadal Oscillation (AMO), a major TP in the Atlantic Ocean. A recent study [12] showed that an increase in sea surface temperatures of the oceanic basins during 1984–2006 was not attributable to an increase in greenhouse gases, but the authors suggested the primary importance of vertical heat transport in the oceans. Donner interpreted flood events at Tarawa in terms of El Niño, low pressure from tropical cyclones, and localized human activities rather than global sea level rise [13]. Johnstone and Mantua [14] reconstructed local surface temperature in northeastern Pacific coastal areas using the Pacific Decadal Oscillation (PDO, a major TP in the Pacific Ocean) and sea-level pressure. They showed thereby the dominant role of natural variability from monthly to century time scales. Wunsch and Heimbach [15] examined thermal changes in the abyssal ocean for 1992-2011 and suggested the existence of long memory (decades to thousands years) of the deep ocean producing trendlike changes in abyssal heat content.

**Cloud;** Stephens et al. [16] thoroughly reviewed the effect of cloud on the albedo, and made following conclusions; (i) the Northern and Southern Hemisphere (NH, SH) reflect the same amount of sunlight within ~ 0.2 W/m$^2$, which symmetry is a result of an increase in SH clouds offsetting the reflection from the NH land masses; (ii) the albedo of Earth is highly buffered on hemispheric and global scales owing to high degree of freedom provided by clouds. Such regulation due to clouds is essentially important in the climate system although, according to their claim, current climate models cannot sufficiently reproduce it. They also pointed out that the regions of highest variability coincided with regions showing greatest tropical moist convection variability and that the interannual variability of the tropic cloudiness was associated with El Niño-Southern Oscillation (ENSO) which is a dominant TP in the tropical region.

**Sun;** it has been shown that solar activity changes are correlated with historical records such as the depth variations of Lake Naivasha in East Africa [17] and variability of the Asian monsoon [18]. Kodera [19] proposed ultraviolet (UV) region radiation-induced changes of atmospheric circulations by considering slight changes of total solar irradiance (TSI) during the 11-year solar cycle (~0.1%). Elsner et al. [20] interpreted the inverse relationship between the number of hurricanes over the Caribbean and sunspot number (SSN) on the basis of daily variations in UV intensity. There have been reports on the relationship between geomagnetic activity and the Arctic Oscillation (AO) and North Atlantic Oscillation (NAO), major TPs in the Northern Hemisphere [21–23], and on proposals for mechanisms of the solar wind–climate connection [e.g., 24, 25]. We have demonstrated that a correlation between surface temperature ($T_s$) and the *aa* index, an effective indicator of solar wind activity [26], is strong in Northern Europe and the Atlantic Ocean during certain seasons [27]. Takahashi et al. [28] and Hong et al. [29] found periodic variations in cloud activity, which they interpreted using the 27-day solar rotational period. Substantial activity changes were observed around the tropical warm pool region during the high solar activity period, and they suggested that the 11-year solar cycle affected large-scale oceanic dipole phenomena.

**Moon;** Cerveny and Shaffer [30] linked El Niño and the Luni-Solar oscillation (LSO) by showing a correlation between lunar declination and the ENSO. Ramos da Silva and Avissar [31] predicted that the AO index would have a positive peak in 2010 on the basis of its relationship with the LSO, and their prediction seems successful. Yasuda [32] showed an 18.6-year period of a moon-tidal cycle in the PDO.

When the above observations of natural factors are verified, one of our scenarios assuming large natural climate variations will not be unrealistic. The influences of the sun and the moon are, however, still controversial, as manifested by discussions in the Intergovernmental Panel on Climate Change (IPCC) Fifth Assessment Report [33]. In fact, there is not a single word on the effect of the moon in this report. The solar influence was not regarded as major. The mechanism of cosmic rays [e.g., 34] was regarded as doubtful and the solar wind was not addressed at all. Only UV variation was considered a plausible factor.

If the IPCC failed to grasp the reality of climatic impacts of the sun and moon, climate policy based on its analysis would not be effective. If these impacts prove strong, they will undoubtedly affect future climate policy. A challenge is therefore to evaluate the magnitudes of the impacts at the regional/local scale in the short term. Thus, as a first step in the present study, we elucidate the influence of changes in solar activity. We pay particular attention to the solar wind for two reasons: 1) the literature cited above appears scientifically sound; 2) the influences of the solar wind and UV can be easily distinguished, because their time dependences are very different.

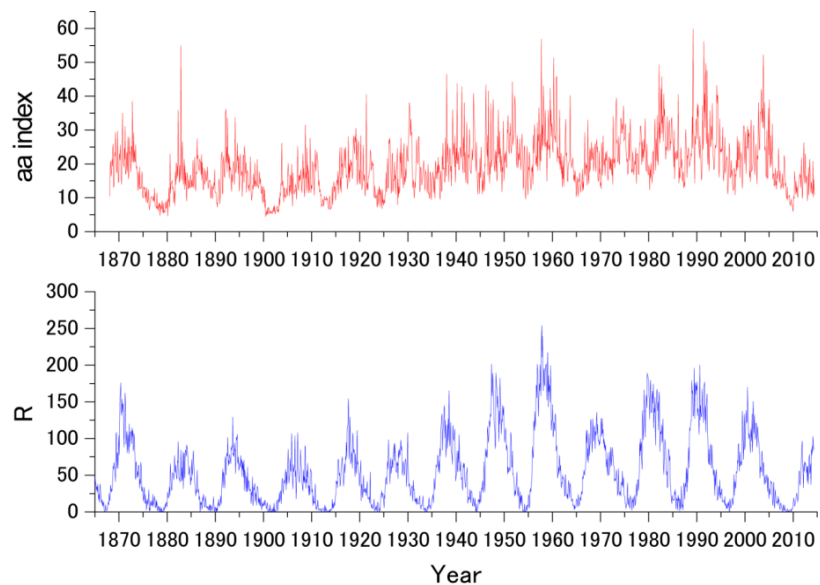

**Fig. 1. Variations of aa index and sunspot number (R) during 1900–2014.**

That is, the UV follows the 11-year cycle but the solar wind does not (**Fig. 1**) because of the contribution of low-latitude coronal holes on the solar surface during the declining period after a sunspot maximum [e.g., 35]. This requires consideration of multiple pathways, given the establishment of the influence of UV change. Both the UV and solar wind can affect the climate.

We therefore scrutinize the solar wind influence to establish the multi-pathway scheme. To do so, we take a phenomenological approach in which we focus on the relationship between solar wind activity and climate indices (temperature and TPs such as the AO). That is, we examine the effect of the solar wind on regional/local monthly temperatures and compare it with that of the TPs. Furthermore, we consider the UV effect by addressing SSN changes. The effect of the moon should also be considered, but this awaits future study.

In standard meteorological analyses, the solar influence is typically assessed using measures such as geopotential heights and mean zonal wind [34–36]. In the present analysis, however, we use monthly or seasonal regional/local surface temperature data. This is mainly because shorter and more regional variations are important to actual environmental policies, and temperature is a suitable index for this purpose.

As an indicator of the solar wind, we use the *aa* index and $P_\alpha$ (work incorporated from the solar wind by the earth magnetosphere), the latter being directly calculated from solar wind data measured at one of the Lagrangian points ($L_1$, 1.5 million km from the earth) [26]. It is important that $P_\alpha$ is free from the influence on *aa* of terrestrial factors, such as waves propagating from the troposphere into the ionosphere and modulating *aa* via changes in electric conductance of the ionosphere [39]. The phase of the quasi-biennial oscillation (QBO) of equatorial stratospheric wind was also taken into account, considering the well-known QBO involvement in the relationship between solar activity (SSN and 10-cm flux) and atmospheric circulation [37].

Using correlation maps (spatial distribution of the correlation coefficient) [27] and correlation coefficients based on time series (interannual variations of particular months) of the climate indices, we found that the solar wind influence is as great as that of the TPs. This naturally leads to the conclusion that the former affects the latter. Furthermore, the influences of the QBO and SSN were demonstrated to be important. These findings verify the importance of the solar wind as a natural effect on regional/local climates.

## 2. Method
### 2.1. Indices for the climate system
#### 2.1.1. Temperature data

Monthly and seasonal gridded data for surface temperature ($T_s$) datasets were obtained from a NASA website (http://data.giss.nasa.gov/gistemp/station_data/). The gridded data consist of temperature anomalies (base period 1951–1980) and include ocean regions; grid size is 2° × 2°.

Monthly gridded temperature data for the stratosphere ($T_{st}$) and troposphere ($T_{tr}$) were obtained from a Remote Sensing Systems website (http://www.remss.com/measurements/upper-air-temperature).

#### 2.1.2. Teleconnection pattern indices

Based on the observation that TPs strongly contribute to natural climate variation as described in the introduction section, we consider major TPs here. TP indices were acquired from the Climate Prediction Center (http://www.cpc.ncep.noaa.gov/).

TPs used were the AO, Antarctic Oscillation (AAO), Southern Oscillation (SO), AMO, PDO, and Pacific North American (PNA) mode. The NAO may be regarded as part of the AO, and El Niño is closely related to the SO; hence, they were not included here although they and other TPs are worth considering in detailed analysis.

### 2.1.3. Quasi-Biennial Oscillation

The QBO phase (westerly or easterly wind) in a given month and year was determined using data from the period 1942–2014 (http://www. esrl.noaa.gov/psd/data/correlation/qbo.data) for stratification in calculating correlation.

## 2.2. Indicators of solar activity
### 2.2.1. *aa* index and SSN

Values of *aa* were obtained from a database of the British Geological Survey (http://www.geomag.bgs.ac.uk/data_service/data/magnetic_indices/aaindex. html).

Monthly data of SSN (denoted as $R$ below) were procured from SILSO data/image, Royal Observatory of Belgium, Brussels (http://www.sidc.be/silso/monthlyssnplot).

### 2.2.2. Solar wind parameters

OMNI2 data, which contains solar wind parameters observed at point $L_1$, were obtained from http://omniweb.gsfc.nasa.gov/ow.html. The OMNI 2 dataset is not continuous, and months with few data were not included in the analysis. Finch and Lockwood [26] excluded months containing more than 10 days of no data. In the present analysis, however, the data were too sparse using that threshold, because we examined month-to-month correlations whereas Finch and Lockwood [26] used the entire dataset.

According to Finch and Lockwood [25], $P_\alpha$ (a theoretical solar wind-magnetosphere coupling function representing power extracted from the solar wind by the magnetosphere) is most suitable to address the relationship between geomagnetic activity (such as the *aa*) and the solar wind. We used $P_\alpha$ because of its clear physical meaning.

$$P_\alpha = (k\pi/2\mu_0^{(1/3+\alpha)})\ m^{(2/3-\alpha)} M_E^{2/3}\ N^{(2/3-\alpha)}\ V^{(7/3-\alpha)}\ |\boldsymbol{B}|^{2\alpha} \sin^4(\theta/2) \qquad (1)$$

Here, k is a constant, $m$ is mean ion mass, $M_E$ is the magnetic moment of the earth (assumed to decrease by about 5% per century), $N$ is solar wind particle density, $V$ is solar wind velocity, $\boldsymbol{B}$ is the solar wind magnetic field, and $\theta$ is the clock angle of the interplanetary magnetic field in the Geocentric Solar Magnetic reference frame (i.e., the angle between the magnetic field of the solar wind and that of the earth). Coupling exponent $\alpha$ was set to 0.3, but dependence of $P_\alpha$ on $\alpha$ was weak.

## 2.3. Correlation coefficients and correlation maps

We examined the correlations for the following combinations: between $T_s$ and the solar wind parameters ($aa$ and $P_\alpha$); between $T_s$ and indices of the TPs (AO, AAO, SO, AMO, and PNA); and between the solar wind parameters and the TP indices. All data were used without preprocessing such as detrending, averaging, and filtering. This is because temporal and spatial variations of the temperature data are large relative to their trends (cf. **Fig. 4**).

Besides the correlation coefficient ($r$) and significance ($p$-value) based on time series of the indices, we used correlation maps [27], i.e., spatial distributions of the correlation coefficient when the combination contained the gridded $T_s$ data. We compared correlation maps for the solar wind parameters and those for the teleconnection indices, which enabled further examination of their relationship. Similarity between the correlation maps was evaluated via normalized cross correlation (NCC, its value being $R_{NCC}$):

$$R_{NCC} = \sum_{i,j} M_1(i,j) M_2(i,j) / \sqrt{\sum_{i,j} M_1(i,j)^2 \sum_{i,j} M_2(i,j)^2} \qquad (2)$$

Here, $M_1(i, j)$ and $M_2(i, j)$ denote matrices for two correlation maps; i (1–180) and j (1–90) correspond to longitude and latitude, respectively.

In calculating $r$, we focused on late winter to early spring months (mainly January for $aa$ and $P_\alpha$, and January through May for $T_s$ and teleconnection pattern indices) of the years. This is because we have already observed a strong correlation between winter $aa$ and spring $T_s$ in certain areas, such as Northern Europe [27]. Similarly, Palamara and Bryant [22] reported that $aa$ is strongly correlated with the AO during winter.

The time window used was from 20 to 73 years. The shortest window was limited by statistical significance (number of points, $n$, was about 10 when stratified by QBO phase), and the longest one was the same as the period during which the QBO phase is reliable.

## 3. Results and Discussion
### 3.1. Relationships between *aa* index and *P*α

First, we show that use of $aa$ as a measure of the solar wind in the present analysis is justified. Although a correlation between $aa$ and $P_\alpha$ is strong for a long period [26], this does not necessarily mean that the correlation is strong for monthly data. For instance, if surface weather conditions have powerful effects on electric conductance of the ionosphere [39], $aa$ may give larger errors than $P_\alpha$ in the correlation analyses. Thus, it is necessary to examine if the use of $aa$ is suitable for our discussion, based on the correlation maps.

**Figure 2** shows an example of the correlation maps of $aa$ vs. $T_s$ and $P_\alpha$ vs. $T_s$, under the same conditions. Because map patterns for the two measures were always very similar, we concluded that *aa* can be used as a proxy measure of $P_\alpha$. We use *aa* in the following analyses because this index has no lack of data during the periods examined.

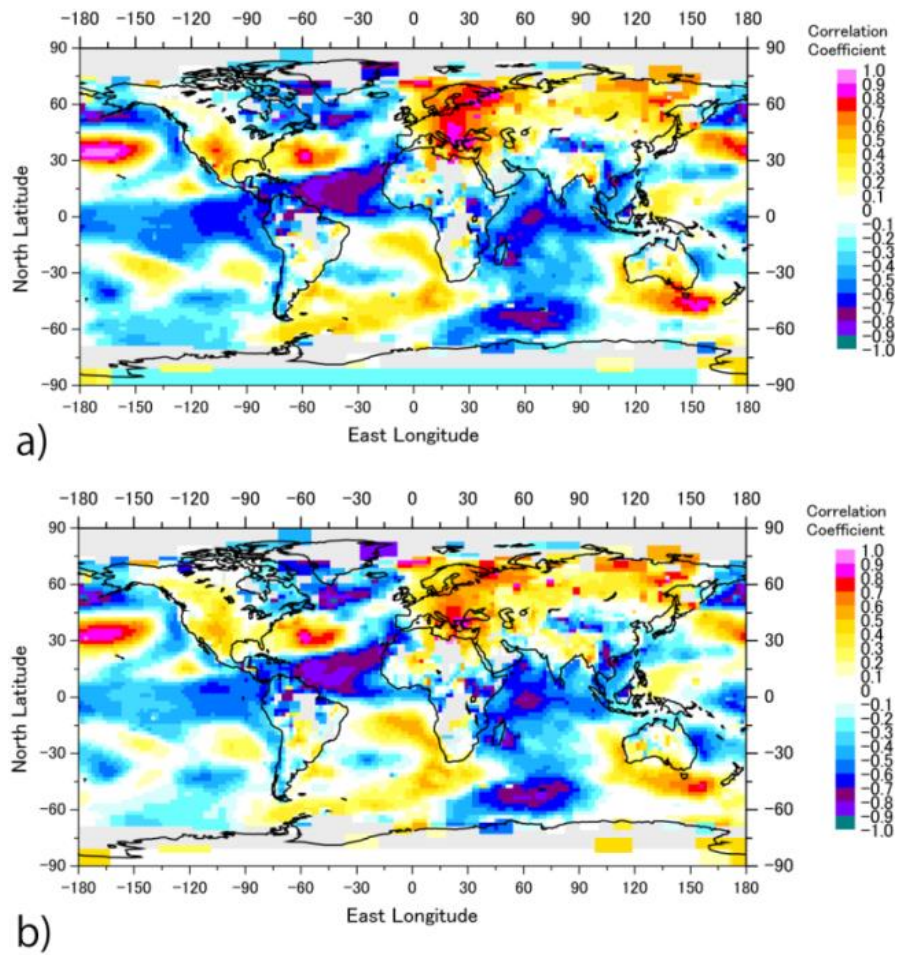

**Fig. 2.** Examples of correlation maps for *aa* and $P_\alpha$ under the same conditions (January for *aa* and $P_\alpha$, March $T_s$, easterly QBO; period 1981–2000). a) for *aa* vs. $T_s$ and b) $P_\alpha$ vs. $T_s$.

### 3.2. Correlation maps of aa vs. $T_s$

**Figures 3** and **4** show correlation maps of *aa* (January) vs. $T_s$ (January–May) for the westerly (**Fig. 3**) and easterly (**Fig. 4**) phases of the QBO. As with *aa*, the month used in the stratification based on the QBO phase was set to January, although other months gave slightly different results. The correlation calculation was carried out with time window of 20 years (19 years for 1942–1960) and with a time shift of 10 years.

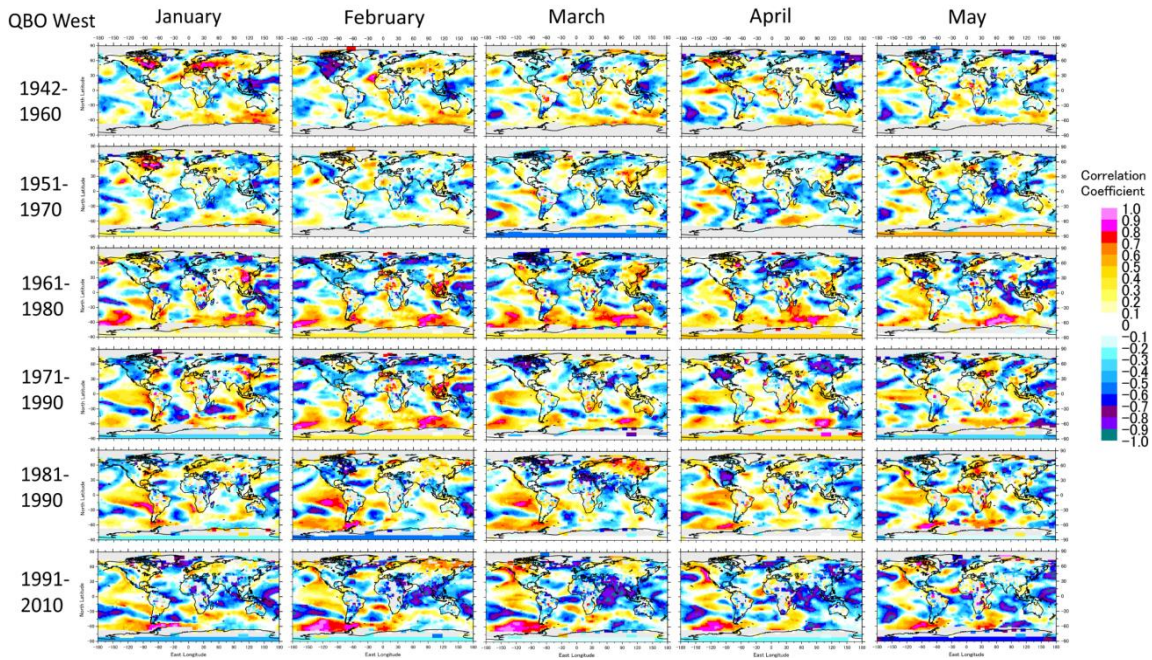

**Fig. 3.** Correlation maps for *aa* (January) vs. $T_s$ (January through May) for the period 1942–2010 with a time window of 20 years and a time shift of 10 years. Phase of QBO is westerly.

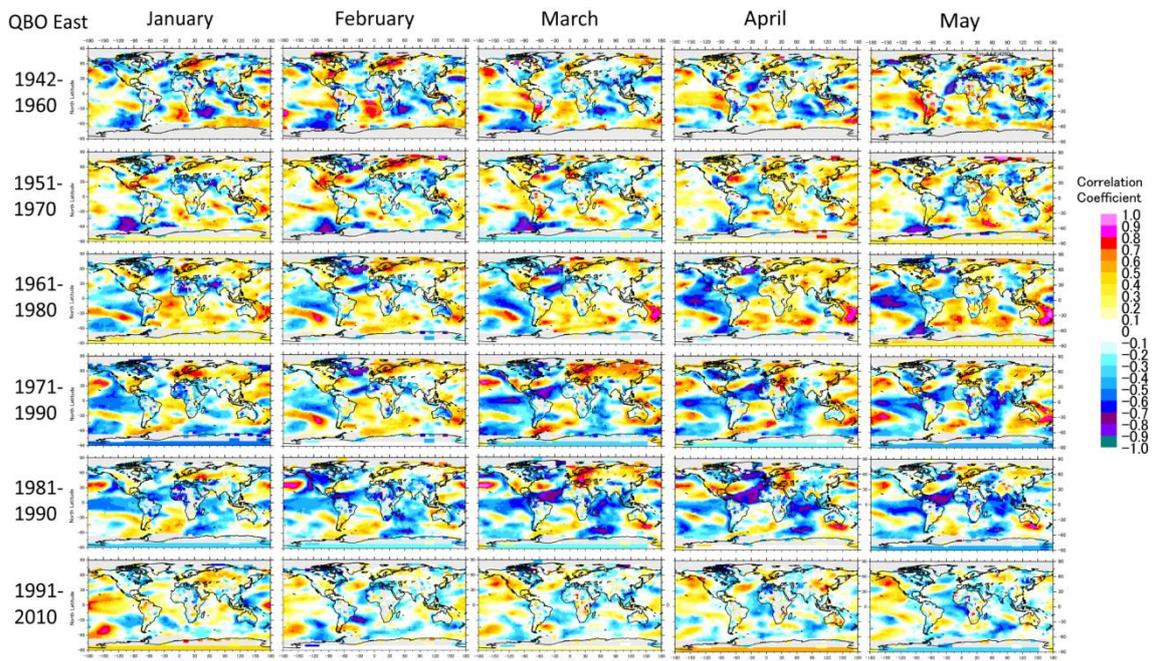

**Fig. 4.** Correlation maps for aa (January) vs. Ts (January through May). Phase of QBO is westerly. Other conditions are the same as in Fig. 2.

In this analysis, *r* values of 0.65–0.7 corresponded to a significance level greater than 95% ($p < 0.05$). Regions with this level were common in the correlation maps; even regions with $r > 0.8$ ($p < 0.01$) were found.

Correlation maps for various months in each 20-year period often revealed common patterns. This is mainly because the $T_s$ spatial pattern tends to persist for several months. For instance, a pattern with positive high correlation was observed around 60°S for 1961–1980 and 1991–2010 with easterly QBO. For 1981–2000, the equatorial region of the Pacific Ocean frequently shows a pattern with negative high correlation in its western portions and a positive correlation in its eastern portions.

### 3.3. "Singular spots" of high correlation

**Figure 3** (westerly QBO) shows "singular spots" of strong correlation throughout the study period. A typical example is a strong negative correlation in the region near 180°E, 15°N (cf. purple circles in **Fig. 5**).

**Figure 5** shows correlation maps and time series for January. The top figures (**A**) are correlation maps of *aa* vs. $T_s$ for different periods, in which the singular spots are shown by purple circles. The bottom figure (**B**) depicts time series at a singular spot; *r* between the two time series is −0.781 and significance is high (number of points n = 32, $p < 0.0001$). Although such singular spots are few, their existence may prove the influence of the solar wind on climate.

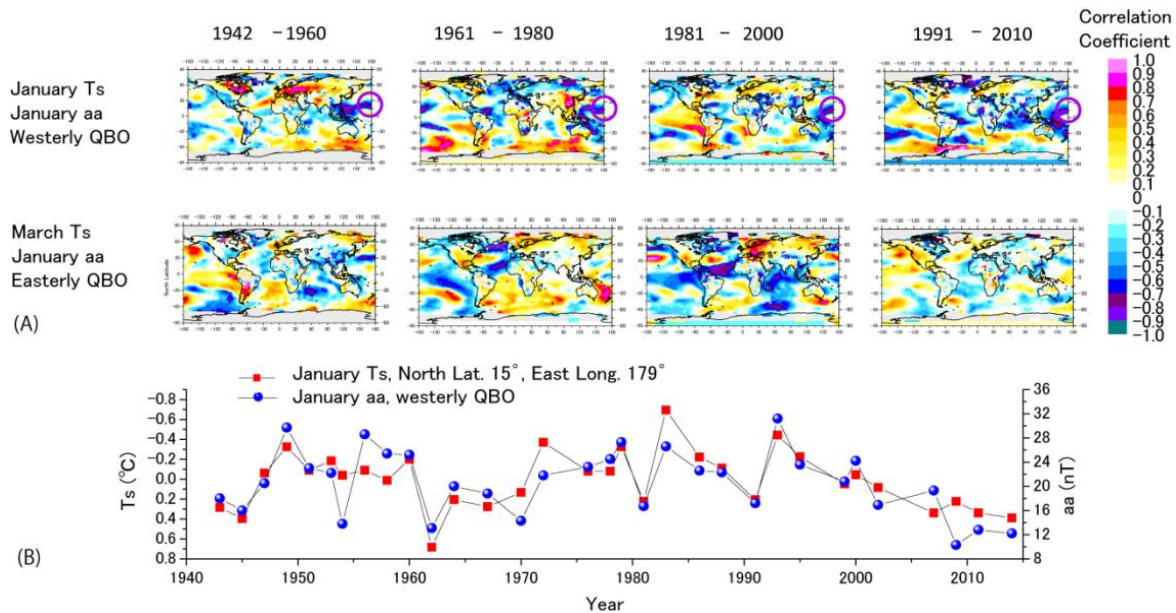

**Fig. 5. Selected correlation maps and time series for *aa* vs. $T_s$. A) Correlation maps for different periods. B) Time series for the period 1942–2014 (red squares for $T_s$, blue disks for *aa* at one of the "singular spots" (15°N, 179°E), purple circles in A).**

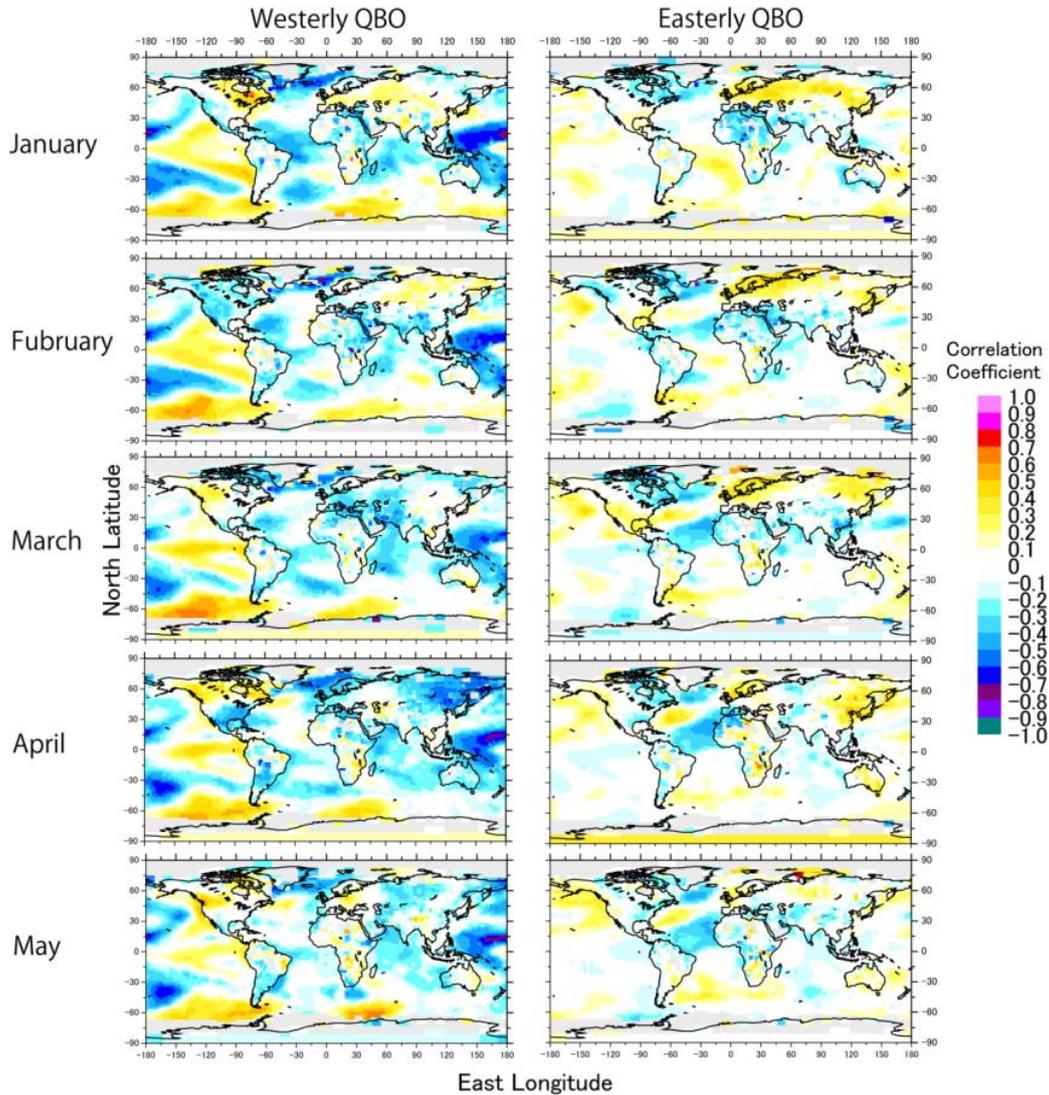

**Fig. 6. Correlation maps for the long period (1942–2014), for various QBO phases and months (January through May).**

**Figure 6** shows correlation maps for 1942–2014, in which singular spots are clearly evident for the westerly QBO (left row) in most months. For the easterly QBO (right row), the spatial pattern resembles that of the AO (cf. **Fig. 7**), but there is no region with high *r*. However, as described below, regions with high *r* appear when the SSN is also considered (cf. **Fig. 10**).

### 3.4. Correlation maps of TPs vs. $T_s$

**Figures 7** and **8** show examples of correlation maps for the combination between TPs and $T_s$. The period of **Fig. 7** is 1942–2014 (1948–2011 for the AAO and 1950–2014 for the PNO), and 20 years for **Fig. 8**. For simplicity, the month for the TPs was the same as that of $T_s$, and the month for the stratification was January, the same as for *aa*.

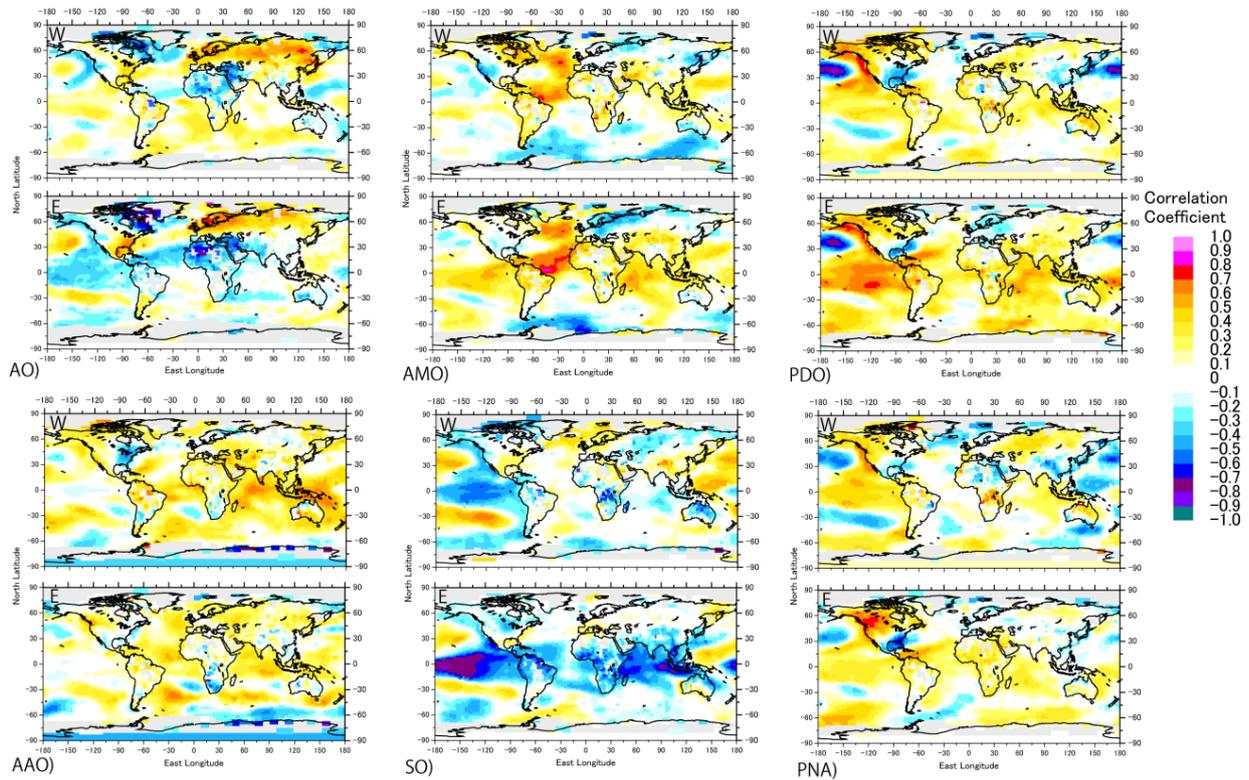

**Fig. 7.** Correlation maps for teleconnection patterns (TP) vs. $T_s$, for 1942–2014 and easterly and westerly QBO phases (1948–2011 for AAO; 1950–2014 for PNO). QBO phase (W or E) is at the upper left in each map, and TP names are at the lower left corner of each pair of QBO phases.

**Figure 7** reveals spatial patterns characteristic to each TP and the influence of the QBO phase. For instance, the correlation map of AO vs. $T_s$ (both January) for 1942–2014 and both westerly and easterly QBO phases shows a positive high correlation in the region from Northern Europe to Siberia and a negative correlation around the Mediterranean Sea, which is a known AO pattern. Over the Pacific Ocean, the easterly and westerly QBO phases have an inverse correlation. Thus, the correlation over the Pacific is weak when the QBO phase is not taken into account.

The correlation maps for the 20-year period in **Fig. 8** show other regions with strong correlations. For the PDO, the characteristic pattern in the equatorial Pacific region is more distinct for the easterly QBO than its westerly counterpart (**Fig. 7**). The pattern in the North Pacific is distinct for the two QBO phases. Long-term characteristic features of the SO and PNA are pronounced for the easterly QBO phase (**Fig. 7**).

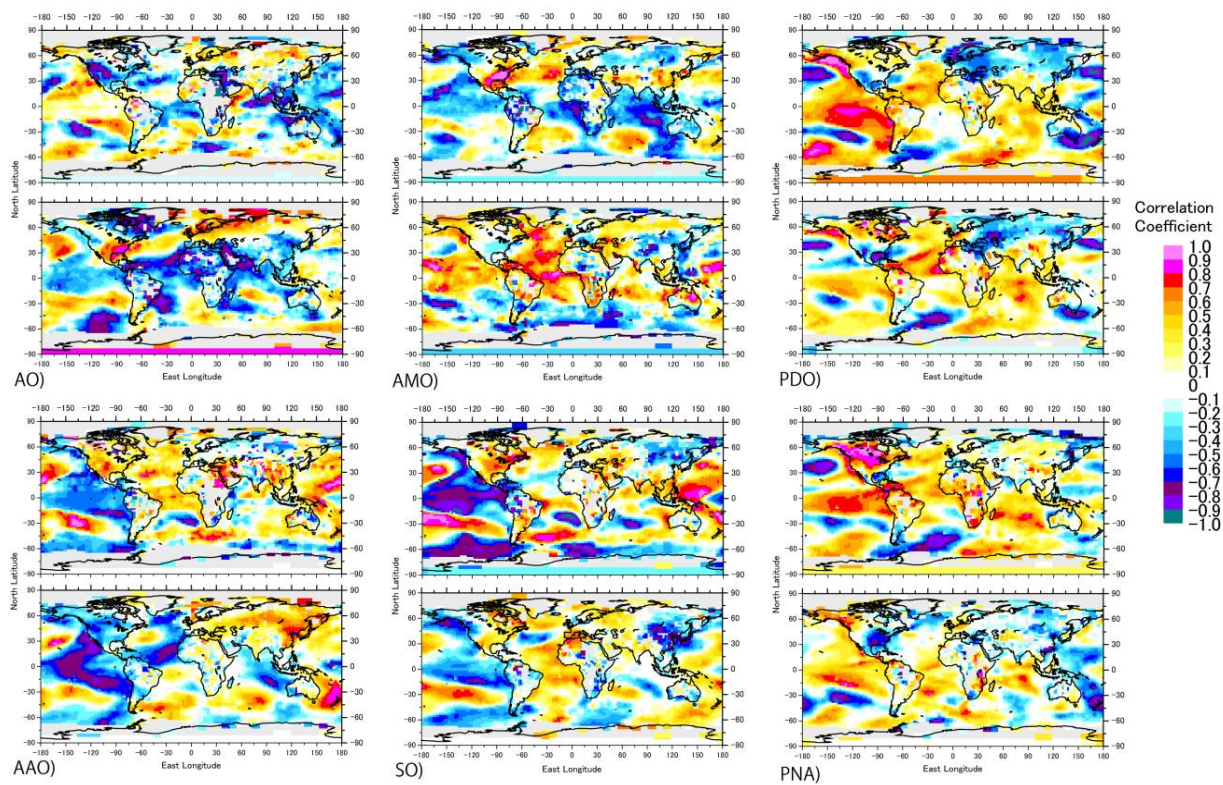

**Fig. 8.** Correlation maps for TPs vs. $T_s$ for the 20-year period. Two examples for each TP with substantially different features are chosen. Details for each map are shown below.

AO: top, January $T_s$ vs. January AO with westerly QBO, 1991-2010: bottom, January $T_s$ vs. January AO with easterly QBO, 1951-1970.

AAO: top, January $T_s$ vs. January AAO with westerly QBO, 1981-2000: bottom, April $T_s$ vs. April AAO with easterly QBO, 1960-1980.

AMO: top, January $T_s$ vs. January AMO with westerly QBO, 1951-1970: bottom, January $T_s$ vs. January AMO with westerly QBO, 1961-1980.

SO: top, February $T_s$ vs. February SO with westerly QBO, 1991-2010: bottom, April $T_s$ vs. April SO, easterly QBO, 1991-2010.

PDO: March $T_s$ vs. March PDO with easterly QBO, 1971-1990: bottom, April $T_s$ vs. April PDO with easterly QBO, 1981-2000.

PNA: top, February $T_s$ vs. February PNA with easterly QBO, 1981-2000: bottom, April $T_s$ vs. April PNA with westerly QBO, 1981-2000.

### 3.4. Relationship between *aa* and TPs

We compared correlations of *aa* vs. $T_s$ and TPs vs. $T_s$ based on the results above. Regarding correlation magnitude, we conclude from visual inspection and global averages of the absolute values of *r* that correlations for *aa* and the TPs have similar magnitudes. This shows that the influence of the solar wind on climate is as powerful as that of the TPs.

We then examined similarity between the correlation maps for *aa* vs. $T_s$ and TPs vs. $T_s$. In long-term correlation maps (**Fig. 7**), there is no TP that shows strong similarity to the solar wind. The singular spots for the solar wind in **Figs. 5** and **6** are not observed in **Fig. 7** for TPs. This is somewhat surprising when considering the supposed relationship between the AO and solar wind [e.g., 22].

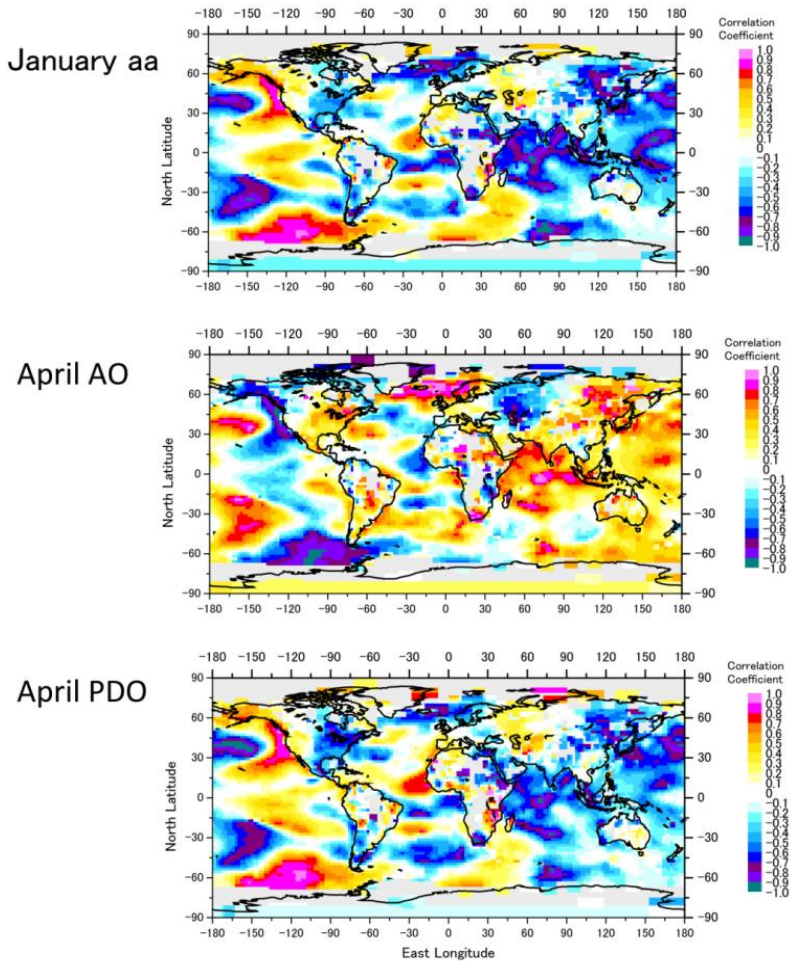

**Fig. 9.** Correlation maps for *aa* vs. $T_s$ (top) and typical TPs vs. $T_s$; AO (middle) and PDO (bottom), for westerly QBO and 1991–2010. Correlation coefficient *r* is as follows: −0.816 (*p* = 0.0135) between *aa* and AO, and 0.926 (*p* < 0.001) between *aa* and PDO.

Strong similarity appears, however, in short-term (20 years) correlation maps, not only for the AO but also for the other TPs (**Fig. 9**). The map for *aa* (**Fig. 9a**) is inversely correlated with that of the AO (**Fig. 9b**) and positively correlated with the PDO (**Fig. 9c**). Pattern similarities between them are very great ($|R_{NCC}| > 0.9$), and *r* values between the time series of their indices are also high ($|r| = 0.8$–$0.9$). Singular spots in **Fig. 9a** for *aa* are also evident in the correlation maps for the TPs (**Fig. 9b** and **c**).

Regarding the AO and PDO, both of which show strong correlations in Fig. 9, *r* values for other periods and months are shown in **Tables 1** (AO) and **2** (PDO). Dark Pink ($r > 0$) and blue ($r < 0$) cells correspond to a significance level greater than 95% ($p < 0.05$). For significance < 95% but $R_{NCC} > 0.6$, cells are shaded light pink and blue.

**Table 1. Correlation coefficient (*r*) for *aa* (January) vs. AO (January through May)**

| QBO | West | | | | | East | | | | |
|---|---|---|---|---|---|---|---|---|---|---|
| | Jan | Feb | Mar | Apr | May | Jan | Feb | Mar | Apr | May |
| 1942-1960 | 0.27 | -0.09 | -0.39 | -0.27 | 0.33 | 0.30 | 0.07 | -0.31 | -0.09 | *-0.65* |
| 1951-1970 | -0.43 | -0.55 | 0.10 | *-0.68* | 0.19 | 0.63 | 0.24 | -0.09 | 0.33 | *-0.61* |
| 1961-1980 | -0.28 | -0.08 | 0.54 | *-0.66* | -0.37 | 0.72 | 0.66 | 0.29 | 0.01 | -0.25 |
| 1971-1990 | -0.19 | -0.24 | 0.21 | *-0.69* | -0.37 | 0.72 | -0.45 | -0.24 | 0.30 | 0.08 |
| 1981-2000 | 0.74 | 0.09 | 0.49 | *-0.70* | *-0.63* | 0.66 | 0.49 | 0.58 | 0.16 | 0.44 |
| 1991-2010 | 0.48 | 0.38 | 0.08 | *-0.82* | *-0.77* | 0.29 | 0.03 | 0.16 | -0.06 | 0.14 |

Dark Pink (r > 0) and blue (r < 0) cells correspond to a significance level greater than 95% (p < 0.05) for correlation between indices. For significance < 95%, cells are shaded light pink and blue when spatial correlation between correlation maps ($R_{NCC}$) > 0.6.

**Table 2. Correlation coefficient (*r*) for *aa* (January) vs. PDO (January through May)**

| QBO | West | | | | | East | | | | |
|---|---|---|---|---|---|---|---|---|---|---|
| | Jan | Feb | Mar | Apr | May | Jan | Feb | Mar | Apr | May |
| 1942-1960 | -0.24 | -0.46 | -0.41 | -0.13 | -0.32 | -0.56 | -0.07 | -0.08 | -0.18 | -0.03 |
| 1951-1970 | -0.16 | -0.05 | -0.36 | 0.02 | 0.27 | -0.17 | 0.15 | 0.12 | -0.40 | 0.35 |
| 1961-1980 | -0.09 | -0.12 | 0.15 | 0.32 | 0.57 | -0.07 | -0.24 | -0.21 | -0.59 | -0.24 |
| 1971-1990 | -0.10 | -0.23 | 0.13 | 0.11 | 0.04 | -0.58 | -0.67 | -0.66 | -0.59 | -0.42 |
| 1981-2000 | 0.17 | 0.09 | 0.36 | 0.45 | 0.49 | -0.58 | -0.76 | -0.78 | -0.77 | -0.49 |
| 1991-2010 | 0.28 | 0.73 | 0.93 | 0.93 | 0.84 | 0.01 | 0.06 | 0.31 | 0.20 | 0.37 |

The meaning of colors is the same as in Table 1.

Cells with high correlations tend to cluster in **Tables 1 and 2**. For the AO, April for westerly QBO and January for easterly QBO have large *r*. For the PDO, the period 1991–2010 for westerly QBO and 1971–1990 and 1981–2000 for easterly QBO have large *r*. A summary including other TPs is shown in **Table 3**.

**Tables 1–3** show strong correlations between *aa* and TPs for many periods and months. There are only three cells without such correlations in **Table 3** (gray cells). The AAO and PNA lack some data for 1942–1960, so *r* could not be calculated. Thus, the number of cells with strong correlations might increase.

**Table 3.** Summary of periods and months of strong correlations between *aa* and teleconnection patterns (TPs)

QBO East

|  | January | February | March | April | May |
|---|---|---|---|---|---|
| 1942-1960 | PDO |  |  |  | AO |
| 1951-1970 | **AO** |  |  |  | AO |
| 1961-1980 | **AO** | **AO**, **AAO** AMO | **AAO**, AMO | **AAO**, AMO PDO, **SO** | AAO, AMO, **SO** |
| 1971-1990 | **AO**, **AAO** PDO, PNA | **AAO**, **PDO** PNA | AAO, AMO **PDO**, SO | AMO, **SO** PDO, PNA | **SO** |
| 1981-2000 | AMO, **AO** PDO | PDO, PNA | AMO, AO PDO | AMO, PDO PNA, SO | AMO, **SO** |
| 1991-2010 |  |  |  | SO |  |

QBO West

|  | January | February | March | April | May |
|---|---|---|---|---|---|
| 1942-1960 |  |  |  | AO |  |
| 1951-1970 |  | AO |  | AAO, **AO** PNA | AAO |
| 1961-1980 | AMO | AMO | AMO, **AO** | AMO, AO | **PDO, PNA** |
| 1971-1990 | SO |  | **AAO** | **AAO**, AO | PNA |
| 1981-2000 | AAO, **AO** SO |  |  | AO, **PNA** | AO |
| 1991-2010 | AAO, SO | **PDO**, **PNA** SO | **PDO**, **PNA** | AO, **PDO** **PNA** | **AAO**, AO **PDO**, **PNA** |

Bold denotes r > 0 and non-bold r < 0. Underline denotes high correlation between indices (r > 0.6, p < 0.05), and otherwise high spatial correlation between correlation maps ($R_{NCC}$) > 0.6.

**Table 3** and **Fig. 9** also show many periods and months with strong correlations between the TPs. Thus, the solar wind may be able to "excite" different TPs at the same time, directly or indirectly. The AO (and possibly AAO) can be directly excited by the solar wind because of interactions between it and the upper atmosphere at high latitudes, for instance, via a mechanism proposed by Seppälä et al. [25] in which planetary wave propagation is modulated by the solar wind. There could be other direct excitation mechanisms, considering that solar wind particles penetrate deep into the atmosphere at locations of weak magnetic field strength [40]. An indirect mechanism may be from interrelations between the TPs. For instance, as Johnstone and Mantua [14] explained, the PDO is largely influenced by the ENSO and Aleutian Low that co-varies with the PNA.

The finding that even the AO, which reportedly has a strong correlation with the solar wind [21, 22], does not constantly show high *r* (**Table 1**), may be related to the nature of the AO/NAO. That is, the NAO does not always have a structure extending into the stratosphere, so is frequently confined to the troposphere [41]. The influence of the solar wind would be significant only with the former configuration.

### 3.5. SSN influence

Because stratospheric ozone is generated by solar UV, the magnitude of UV variations directly affects the ozone layer. The resultant influence on climate is examined in detail by, for example, *Kodera* [19]. We therefore consider the influence of UV variation together with the solar wind. This contributes to the multi-pathway scheme as put forth in the introduction section. The SSN (its magnitude denoted as *R* below) is suitable for this purpose, because the UV variation is regarded as proportional to *R*. In addition, *R* is also relevant to the cosmic ray mechanism [28, 29, 31, 40, 41].

We divided the 365 months during 1942–2014 (January through May) into three categories (each containing 121 or 122 months) using the magnitude of *R*, i.e., large (0.2–28.9), middle (29.1–86.2), and small (86.5–217.4). Because the QBO phase was also considered here, the number of points for each correlation map was ~12, where $r = 0.6$ corresponds to a significance level about 95%.

The *R* categories gave correlation maps with substantially different features, clearly showing the influence of *R*. Strong negative correlation in the central Pacific Ocean (singular spots), which is observed for the westerly QBO over the entire period (1942–2014) in **Figs. 5** and **6**, is more clearly seen for large and middle *R* in **Fig. 10**. A positive correlation in northern Eurasia (a characteristic feature of the AO, cf. **Fig. 7**) is distinct for large *R* and the easterly QBO. The pattern for small *R* and easterly QBO appears similar to that of the SO (cf. **Fig. 7**). Thus, the combination of the solar wind and SSN, or multi-pathway scheme, proved effective in explaining the influence of solar activity variations on surface temperatures.

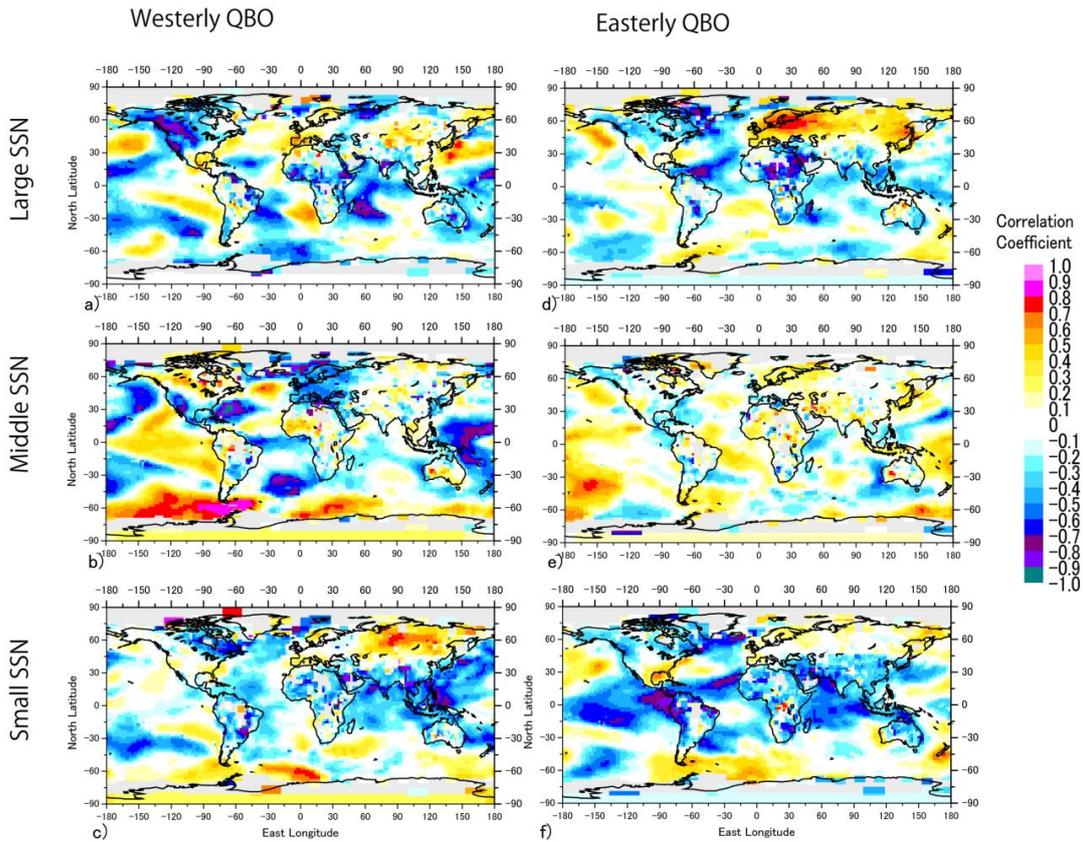

**Fig. 10. Correlation maps obtained by stratification using SSN and QBO phases: February $T_s$ vs. January *aa*. The period examined is 1942-2014 (72 years).**

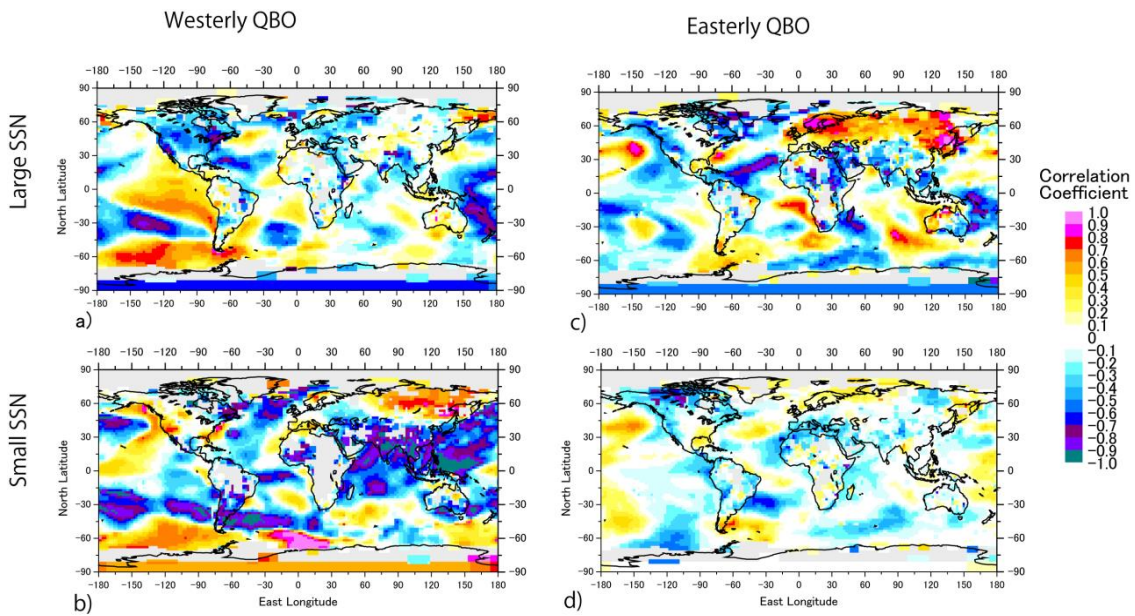

**Fig. 11. Correlation maps obtained by stratification using SSN and QBO phases for February $T_s$ vs. January *aa*. a) westerly QBO, large SSN; b) westerly QBO, small SSN; c) easterly QBO, large SSN; d) easterly QBO, small SSN. The period examined is 1971-2010 (40 years).**

The consideration of *R* is interesting for shorter periods, although there is a statistical limit imposed by the number of points. For instance, **Fig. 11** shows the case of 1971–2010 (40 years), for which the months were divided into two categories by *R* and there were 10–12 points for each correlation map. The characteristic feature in **Fig. 8** is also observed in **Fig. 9**. The singular spots of the westerly QBO phase are seen for large *R*. Further analyses may demand more suitable methods, which are a future challenge.

### 3.6. Temperatures of stratosphere and troposphere

Similar correlation maps were produced for temperature of the stratosphere and troposphere, although satellite observation is limited to the period after 1979. **Figure12** shows typical examples for December *aa* and westerly QBO. We see that the stratospheric pattern is nearly opposite that of the troposphere. The former shows band-like shapes of positive correlation in the equatorial region, and there is a structure of wave number four in the Southern Hemisphere, possibly caused by a planetary wave. The pattern in the troposphere is closer to that of the surface than that of the stratosphere. For example, a feature characteristic of the AO (cf. **Fig. 5**) is evident.

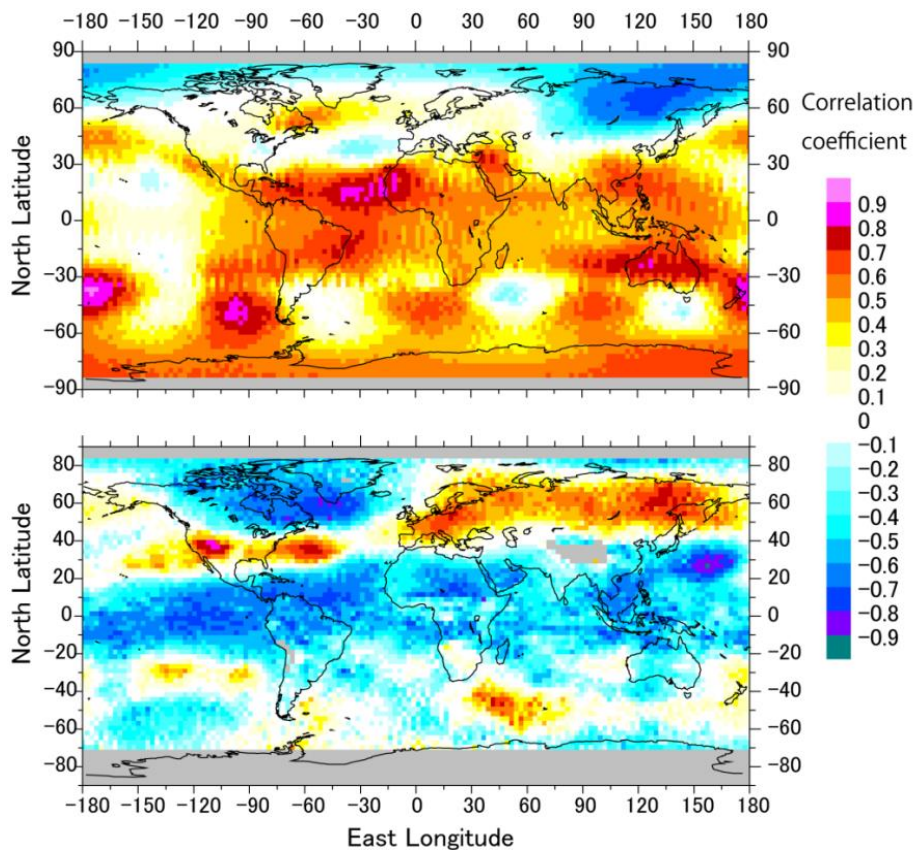

**Fig. 12. Examples of correlations between *aa* and temperatures of stratosphere and troposphere.**

The influence of the solar wind must penetrate the stratosphere and troposphere before reaching the surface. Therefore, detailed examination of the stratosphere and troposphere is necessary in further analyses.

## 4. Conclusions

In this work, we examined the influence of the solar wind on climate, using maps of correlation between solar wind activity and temperature data (surface temperature in particular). The following are the results: 1) the magnitude of the solar wind influence on climate compares well with that of TPs such as the AO and PDO; 2) the solar wind appears to excite the TPs, thereby affecting surface temperature; 3) the manifestation of the solar wind effect largely depends on the period and conditions used; 4) stratification by QBO phase, which is already known, is inevitable; 5) SSN (a measure of solar UV) is also an important factor; 6) the solar wind influence was clearer for short periods (20 years) than for a long one (1942–2014), but there were singular spots where the influence persisted through the long period.

Thus, the influence of the solar wind on climate should be considered much stronger than conventionally believed. Once its mechanism is elucidated and incorporated into climate models, it will greatly contribute to policy development. In other words, the effectiveness of climate models is greatly reduced when the influence of the sun (and moon) is not adequately represented.

Considering the present observation that a single TP such as the AO cannot explain all features of the correlation maps for the solar wind, it is important to examine under which conditions TPs are excited by the solar wind. It is interesting to develop suitable combinations of TPs to represent the influence of the *aa* index on climate. For more productive discussions, detailed studies are necessary on the influence of the solar wind on the thermosphere and stratosphere, as well as on processes by which that influence propagates to the troposphere [24, 25]. This demands cooperation between researchers across wide disciplines. Although there are numerous future challenges, the correlation map method described herein will be a powerful tool for resolving those challenges.

## Acknowledgments

We appreciate the discussions at the Workshop on Solar Activity and Climate (Nagoya, Japan, November 15, 2010). This work was supported by KAKENHI (23651009), a Grant-in-Aid for Challenging Exploratory Research of the Japan Society for the Promotion of Science.